\documentclass[aps,prc,preprint,amsmath,amssymb,showpacs,preprintnumbers,superscriptaddress]{revtex4-1}
\usepackage{CJK}
\usepackage{graphicx}% Include figure files
\usepackage{dcolumn}% Align table columns on decimal point
\usepackage{bm}% bold math
\usepackage{color}
\usepackage{hyperref}% add hypertext capabilities
\usepackage{ulem}%set underline
\allowdisplaybreaks[4]

\begin{document}
\begin{CJK*}{UTF8}{}

\title{An efficient solution for Dirac equation in 3D lattice space with the conjugate gradient method}
\author{B. Li \CJKfamily{gbsn} (李博)}
\affiliation{State Key Laboratory of Nuclear Physics and Technology, School of Physics, Peking University, Beijing 100871, China}

\author{Z. X. Ren \CJKfamily{gbsn} (任政学) }
\affiliation{State Key Laboratory of Nuclear Physics and Technology, School of Physics, Peking University, Beijing 100871, China}

\author{P. W. Zhao \CJKfamily{gbsn} (赵鹏巍)}
\email{pwzhao@pku.edu.cn}
\affiliation{State Key Laboratory of Nuclear Physics and Technology, School of Physics, Peking University, Beijing 100871, China}

\begin{abstract}
An efficient method, preconditioned conjugate gradient method with a filtering function (PCG-F), is proposed for solving iteratively the Dirac equation in 3D lattice space for nuclear systems.
The filtering function is adopted to avoid the variational collapsed problem and a momentum-dependent preconditioner is introduced to promote the efficiency of the iteration.
The PCG-F method is demonstrated in solving the Dirac equation with given spherical and deformed Woods-Saxon potentials.
The solutions given by the inverse Hamiltonian method in 3D lattice space and the shooting method in radial coordinate space are reproduced with a high accuracy.
In comparison with the existing inverse Hamiltonian method, the present PCG-F method is much faster in the convergence of the iteration, in particular for deformed potentials.
It may also provide a promising way to solve the relativistic Hartree-Bogoliubov equation iteratively in the future.
\end{abstract}

%\pacs{21.60.Jz, 21.10.-k, 21.10.Re, 27.20.+n}
% 21.60.Jz Nuclear Density Functional Theory and extensions
%21.10.-k Properties of nuclei; nuclear energy levels
%21.10.Re Collective levels
%21.60.Ev Collective models
%21.60.Cs   Shell model
%23.20.-g Electromagnetic transitions
%23.20.Js Multipole matrix element
%27.20.+n  6 A 19
%27.60.+j 90  A 149
%27.50.+e 59  A  89

\maketitle

\end{CJK*}

%*********************************************************%
%---------------------Introduction------------------------%
%*********************************************************%
\section{Introduction}
During the past decades, new experimental facilities with radioactive beams have extended our knowledge of nuclear physics from the stable to the unstable nuclei far from the stability line.
The density functional theory (DFT) has been proved to be an important microscopic approach for self-consistent description of nuclei~\cite{Bender2003Self}.
Starting from a universal energy density functional, DFT can provide a satisfactory description for nuclei all over the nuclide chart.
The covariant density functional theory (CDFT), which includes the Lorentz symmetry, has attracted a lot of attention in nuclear physics~\cite{meng2016relativistic,RING1996PPNP, meng2006PPNP,Vretenar2005PhysicsReport,NIKSIC2011PPNP}.
Its starting point is a standard effective Lagrangian density,
where nucleons can be coupled with either finite-range meson fields \cite{Long2004PK1, lalazissis2005new}
or zero-range point-coupling interactions \cite{Niksic2008DDPC1, ZhaoPC-PK1}.
The CDFT brings many advantages to describe the nuclear systems, such as the natural inclusion of spin-orbit interactions~\cite{Sharma1993Phys.Lett.B9} and the self-consistent treatment of the time-odd fields~\cite{Meng2013FT_TAC, Zhao2018Spectroscopies}.

An essential ingredient of DFT is to solve the so-called Kohn-Sham equation for nucleons.
In many cases, it has been solved by an expansion of a finite set of basis functions, such as, the eigenfunctions of a harmonic oscillator.
This method has been used successfully for many investigations in the literature, but it has limitations:
(a) The convergence of the number of basis functions depends on the parameters of the basis, and this requires a careful optimization of these parameters, for example, in describing nuclear states with large deformations;
(b) For heavy nuclei, the required number of basis functions becomes large, and the construction and diagonalization of the single-particle Hamiltonian matrix with a large dimension in each step of the iteration leads to a surge of computational costs;
(c) There are specific difficulties in describing nuclei with a large space distribution, for instance, in the case of halo nuclei~\cite{Tanihata1985halo,meng2006PPNP, meng2015halos}.
Therefore, the methods developed in coordinate space are preferred to avoid these limitations.

The CDFT has been successfully developed in spherical coordinate space by solving the spherical relativistic Hartree-Bogoliubov (RHB) equation, where the conventional shooting method works quite well~\cite{meng1998NPA}. It has been applied to investigate the halo and giant halo phenomena in nuclei~\cite{meng1996relativistic, meng1998giant}.
Recently, this framework has also been used to explore the limits of the nuclear landscape~\cite{xia2018ADNDT}.
For deformed nuclei, however, the shooting method becomes rather complicated due to the difficulty in solving the coupled channel differential equations~\cite{Price1987Self-consistent}.
Therefore, the Dirac Woods-Saxon (DWS) basis was proposed~\cite{zhou2003spherical} and the corresponding DWS basis expansion method 
has been used in the CDFT for deformed halo in nuclei~\cite{zhou2010neutron, li2012deformed, SUN2018C22, zhang2019forbidden}.
The same basis has also been employed in the development of spherical and axially deformed relativistic Hartree-Fock theories~\cite{Long2010RHFB, Long2010RHFB_halo_pss,geng2020defRHF}, where nonlocal potentials are involved in.   
Despite these achievements, the CDFT calculations with further inclusions of triaxial and/or octupole deformations become very sophisticated in the DWS expansion method. Therefore, to develop the CDFT in three-dimensional (3D) lattice space without any symmetry limitation is highly desired. 

The nonrelativistic DFT in 3D lattice space has been realized for a long time with the iterative methods including the imaginary time method~\cite{davies1980imaginary, bonche2005solution} and the damped-gradient method~\cite{Reinhard1982comparative, Maruhn2014CPC, Wang2018Nilsson3D}.
The basic idea of these iterative methods is searching for the descending direction of energy and following it iteratively until a local energy minimum is reached.
For covariant DFT, however, due to the existence of Dirac sea in Dirac equation, the relativistic ground state within the Fermi sea is a saddle point rather than a minimum.
A direct application of the iterative methods for Dirac equation usually encounters the so-called variational collapse problem~\cite{zhang2009first, ZhangIJMPE2010}.
Therefore, the development of the CDFT on a 3D lattice becomes a longstanding challenge.
Recently, the inverse Hamiltonian method (IHM) has been used to avoid the variational collapse problem~\cite{hagino2010iterative}, and to solve the CDFT in 3D lattice space~\cite{tanimura20153d}.
Later on, the Fourier spectral method has been used to solve the fermion doubling problem~\cite{REN2017Dirac3D}, another challenge in the numerical implementation of the IHM in lattice space.
This new framework is then successfully applied to study the nuclear linear-chain~\cite{Ren2019C12LCS} and toroidal structures~\cite{REN2020Toroidal}.
Very Recently, the time-dependent CDFT has also been developed on a 3D lattice and applied to investigate the microscopic dynamics of the linear-chain cluster states~\cite{Ren2020HeBeTDCDFT}.

Although the IHM has been successful to solve the Dirac equations in 3D lattice space, it is still very time-consuming for heavy nuclei, due to the numerical complexity for calculating the inverse of the Dirac Hamiltonian.
Moreover, it is practically difficult to apply the IHM to Hartree-Fock-Bogoliubov (HFB) calculations in 3D lattice space due to the slow convergence in calculating the inverse of a HFB Hamiltonian~\cite{Tanimura2014PhdThesis}, although it is feasible in the spherical case~\cite{Tanimura2013HFB_IHM}. Therefore, it is desirable to develop an efficient method to solve the Dirac equation in 3D lattice space.

In this paper, inspired by the successful application of the conjugate gradient method with a filtering step to solve the Dirac equation for electron systems~\cite{LIN2013LOBPCG-F}, a preconditioned conjugate gradient method with a filtering function (PCG-F) is developed to solve the Dirac equation for nuclear systems on a 3D lattice.
This new method avoids the inverse of Hamiltonian and, thus, provides an efficient way to solve the nuclear Dirac equation in 3D lattice space.
Moreover, it also paves a new way to solve the RHB equation with the powerful gradient method~\cite{ring2004nuclear}.
The efficiency and accuracy of the newly proposed PCG-F method are demonstrated in comparison with the inverse Hamiltonian and shooting methods.

%*********************************************************%
%--------------Theoretical framework----------------------%
%*********************************************************%
\section{Theoretical framework}\label{sec2}
\subsection{Conjugate gradient method for eigenstate problems}
The conjugate gradient method was proposed to solve the system of linear equations iteratively~\cite{hestenes1952CG}.
Later on, it was extended to solve the eigenstate problem~\cite{bradbury1966CGeigen},
\begin{equation}
  A\phi=\lambda \phi,
\end{equation}
where $A$ is a real $n\times n$ symmetric matrix and $\phi$ is the eigenstate with the eigenvalue $\lambda$.
In the conjugate gradient method, the eigenstate with the smallest eigenvalue is obtained by minimizing the Rayleigh quotient,
\begin{equation}
  \min_{X\in\mathbb{R}^n}\lambda(X)=\min_{X\in\mathbb{R}^n}\frac{(X,AX)}{(X,X)}.
\end{equation}
Here, the trial solution $X$ is updated iteratively starting from an normalized initial guess $X^{(0)}$. In the $i$-th iteration, the search direction $P^{(i)}$ for updating $X^{(i)}$ is determined by
\begin{equation}
    P^{(i)}=R^{(i)}+\beta^{(i-1)}P^{(i-1)},\quad \beta^{(i-1)}=-\frac{(R^{(i)},AP^{(i-1)})}{(P^{(i-1)},AP^{(i-1)})}
\end{equation}
with the residual $R^{(i)}=AX^{(i)}-(X^{(i)},AX^{(i)})X^{(i)}$ and $P^{(0)}=R^{(0)}$.
As a result, the updated $X^{(i+1)}$ is provided by
\begin{equation}
   X^{(i+1)}=G^aX^{(i)}+G^bP^{(i)},
\end{equation}
where the coefficients $G^a$ and $G^b$ are chosen to minimize $\lambda^{(i+1)}=(X^{(i+1)},AX^{(i+1)})$ under the normalization condition $(X^{(i+1)},X^{(i+1)})=1$.

\subsection{Locally optimal block preconditioned conjugate gradient method}
For a long time, the conjugate gradient method suffers from a poor convergence in the iteration process of finding the eigenstate.
Therefore, the preconditioning technique has been introduced, and this provides the preconditioned conjugate gradient (PCG) methods~\cite{Knyazev1998Preconditioned}.
Compared to other types of PCG methods, the so-called locally optimal block PCG method~\cite{Knyazev2001LOBPCG}, where the local optimization of a three-term recurrence is adopted, has been shown to be effective for evaluating a relatively large number of eigenvalues and eigenstates.

In the locally optimal block PCG method, the $n$ lowest eigenstates of a Hamiltonian $\hat{h}$ are solved iteratively starting from a sets of normalized guess solutions $X_k^{(0)}~(k=1,2,...,n)$.
The trial wavefunction $X_k$ is updated iteratively with
\begin{equation}
  X_k^{(i+1)}=\sum_{l=1}^n\left(G^a_{kl}X_l^{(i)}+G^b_{kl}R_l^{(i)}+G^c_{kl}P_l^{(i)}\right).
\end{equation}
Here, $R_l^{(i)}$ is the residual
\begin{equation}
  R_l^{(i)}=\hat{h}X_l^{(i)}-\langle X_l^{(i)}|\hat{h}|X_l^{(i)}\rangle X_l^{(i)},
\end{equation}
and $P_l^{(i)}$ is the previous search direction
\begin{equation}
  P_l^{(i)}=X_l^{(i)}-\sum_{l'=1}^n\langle X_{l'}^{(i-1)}|X_l^{(i)}\rangle X_{l'}^{(i-1)}, \quad P^{(0)}=0.
\end{equation}
To accelerate the convergence of the evolution, the preconditioning technique is usually adopted for $R^{(i)}$
\begin{equation}\label{Eq_preconditonedR}
  W_l^{(i)}=\hat{T}_l^{-1}R_l^{(i)},
\end{equation}
where $\hat{T}_l$ is the preconditioner.
As a result, the wavefunction $X_k$ is updated with
\begin{equation}\label{eq_Xip1}
  X_k^{(i+1)}=\sum_{l=1}^n\left(G^a_{kl}X_l^{(i)}+G^b_{kl}W_l^{(i)}+G^c_{kl}P_l^{(i)}\right),
\end{equation}
where the coefficient matrices $G^a$, $G^b$ and $G^c$ are chosen to minimize $\sum\limits_{k=1}^n\langle X_k^{(i+1)}|\hat{h}|X_k^{(i+1)}\rangle$ under the orthonormalization condition $\langle X_k^{(i+1)}|X_l^{(i+1)}\rangle = \delta_{kl}$.

\subsection{Filtering and preconditioning operators for Dirac equation}
The main task for CDFT is to solve the Dirac equation with the Hamiltonian,
\begin{equation}\label{Eq_dirac}
  \hat{h}=\bm{\alpha}\cdot\hat{\bm{p}}+\beta(m_N+S)+V - m_N,
\end{equation}
where $\bm{\alpha}$ and $\beta$ are the Dirac matrices, $m_N$ is the mass of nucleon, and $S$ and $V$ are the scalar and vector potentials, respectively.
For the sake of convenience, here the Hamiltonian is shifted down by a nucleon mass $m_N$.
Since the spectrum of the Dirac Hamiltonian $\hat{h}$ contains negative- and positive-energy states, a direct application of the PCG method would suffer from the variational collapse problem.

To avoid the variational collapse problem, a filtering operator can be used to suppress the components of negative-energy states in the wavefunctions during the iteration~\cite{LIN2013LOBPCG-F}. In the present work, the filtering operator is taken as,
\begin{equation}\label{eq_filter}
  F(\hat{h})=\frac{1}{D^2}(\hat{h}-C)^2,
\end{equation}
where $C$ and $D$ are two parameters to be optimized in the practical calculations.
The filtering operator is implemented in the PCG method by replacing $X_l^{(i)}$ and $W^{(i)}_l$ in Eq.~\eqref{eq_Xip1} with
\begin{equation}\label{eq_filteredXW}
   X_l^{(i)}\rightarrow F(\hat{h})X_l^{(i)}, \quad W^{(i)}_l \rightarrow [F(\hat{h})]^{N_F}W_l^{(i)}.
\end{equation}
Note that here the filtering operation on $W^{(i)}_l$ is carried out by $N_F$ times.
This is different from the solution of Dirac equation for electron systems, where the states in the Fermi- and Dirac sea are well separated due to the negligible spin-orbit splittings. 
For nuclear systems, however, due to the large spin-orbit interactions, the energy gap between the positive- and negative-energy states is only two or three times of the potentials $S$ and $V$.
The optimized value of $N_F$ will be discussed below in Sec.~\ref{Sec_numer}.

The multiple filtering operations on $W^{(i)}_l$ could lead to a poor convergence behavior because the components of high positive-energy states can be substantially enlarged.
Motivated by the fact that the high-energy states are usually dominated by the kinetic energy, here the preconditioner $\hat{T}$ in Eq.~\eqref{Eq_preconditonedR} is chosen as the following momentum-dependent form,
\begin{equation}\label{eq_preconditoner}
  \hat{T}_l=[\hat{\bm{p}}^2+g_l^2m_N^2]^2,
\end{equation}
where $g_l$ is a optimized factor and will be discussed below in Sec.~\ref{Sec_numer}. This preconditioner operation could effectively  damp the components of high-energy states, and provides an efficient convergence.
Note that for electron systems, such a preconditioner operation is not mandatory~\cite{LIN2013LOBPCG-F}.

\section{Numerical details}\label{Sec_numer}
In the present work, the Dirac equation for nucleons is solved in 3D lattice space by the PCG-F method.
The large scalar and vector potentials are taken as a Woods-Saxon form,
\begin{equation}\label{Eq_potential}
  \begin{split}
    &V(\bm{r})+S(\bm{r})=\frac{V_0}{1+\exp[(r-R_0D(\theta,\varphi))/a]},\\
    &V(\bm{r})-S(\bm{r})=\frac{-\lambda V_0}{1+\exp[(r-R_{ls}D(\theta,\varphi))/a_{ls}]},
  \end{split}
\end{equation}
where $D(\theta,\varphi)$ brings in the quadrupole ($\beta$, $\gamma$) and octupole deformations $\beta_{30}$,
\begin{equation}\label{Eq_deformparameter}
  \begin{split}
   D(\theta,\varphi)=&1+\beta\cos\gamma Y_{20}(\theta,\varphi)+\frac{1}{\sqrt{2}}\beta\sin\gamma[Y_{22}(\theta,\varphi)\\
   &+Y_{2(-2)}(\theta,\varphi)]+\beta_{30}Y_{30}(\theta,\varphi).
  \end{split}
\end{equation}
The parameters for the Woods-Saxon potentials are listed in Table~\ref{Tab1}, which correspond to the neutron potentials
of $^{48}$Ca~\cite{koepf1991WoodsSaxon}.

\begin{table}[!htbp]
\caption{\label{Tab1}The parameters in the Woods-Saxon potentials [see Eq.~\eqref{Eq_potential}].}
\begin{ruledtabular}
\begin{tabular}{cccccc}
$V_0$ [MeV]     &$R_0$ [fm]     &$a$ [fm]    &$\lambda$     &$R_{ls}$ [fm]      &$a_{ls}$ [fm]\\
\hline
-65.796        &4.482         &0.615      &11.118        &4.159             &0.648
\end{tabular}
\end{ruledtabular}
\end{table}

In the present calculations, the coordinate space along the $x$, $y$, and $z$ axes is respectively discretized by 28 grids with the mesh size $d=1$ fm.
The initial guess of the single-particle wavefunctions are taken as the   spherical harmonic oscillator wavefunctions for both upper and lower components.
To avoid the fermion doubling problem, the spatial derivatives are performed in the momentum space with the help of the fast Fourier transformation~\cite{REN2017Dirac3D}.

For the filtering operation, we define $C = -2m_N$ in $F(\hat{h})$, and this is in analogy to Ref.~\cite{LIN2013LOBPCG-F}, where $C = -2m_e$ is used for electron systems.
It should be noted that the value of $D$ is in principle irrelative, because it is anyhow absorbed in the coefficient matrices $G^a$, $G^b$, $G^c$ in Eq.~(\ref{eq_Xip1}) via the energy minimization and orthonormalization condition.
In the present work, we define $D = (V+S)_{\rm min}+2m_N$ with $(V+S)_{\rm min}$ being the minimum of the potential $V+S$ in the coordinate space. By this definition, we have $F(\hat{h}) \approx 1$ for the lowest positive-energy state, and $F(\hat{h}) \approx 0.1$ for the highest negative-energy state.

\begin{figure}[!htbp]
  \centering
  \includegraphics[width=0.45\textwidth]{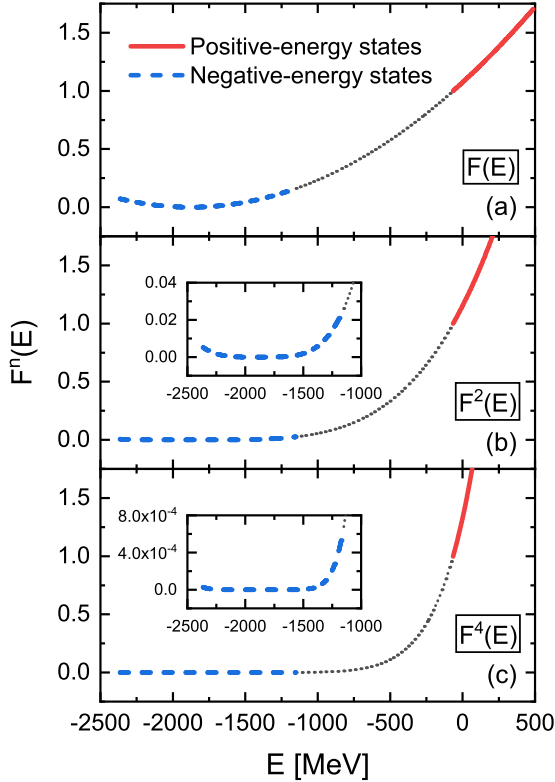}\\
  \caption{(Color online) The filtering function $F^{n}(E)$ with $n=1$ (top), $n=2$ (middle), and $n=4$ (bottom) as a function of energy $E$.
  The solid and dashed lines respectively correspond to the positive- and negative-energy states, which are separated by the filtering function values (dotted lines). The insets present a partial enlargement for the negative-energy region.
  }\label{fig1}
\end{figure}

To illustrate the effects of the filtering operation, in Fig.~\ref{fig1}(a), the filtering function $F(E)$ is shown as a function of energy $E$.
It can be seen that the $F(E)$ values are very small (around $10^{-1}$) in the region of negative-energy spectrum, in comparison with those in the positive-energy spectrum.
This could suppress the negative-energy components during the iteration, while the suppression is found to be not sufficient to avoid the variational collapse in the practical calculations.
Therefore, as in Eq.~\eqref{eq_filteredXW}, the filtering operation on $W^{(i)}_l$ is performed by $N_F$ times.

In Figs.~\ref{fig1}(b) and \ref{fig1}(c), the square and fourth power of the filtering function $F(E)$ are shown respectively.
The suppression of the filtering function on negative-energy states is promoted obviously with the increasing power.
In particular for $F^4(E)$, the suppression on the negative-energy states reaches to an oder of $10^{-4}$.
It is found that such a suppression works quite well to overcome the variational collapse problem and, thus, we adopt $N_F = 4$ in the present work.

In Fig.~\ref{fig1}, one can also see that the filtering function at high energies becomes larger with the increasing power.
This could lead to a slow convergence of the iteration, because the components of bound states are relatively reduced by the filtering function.
Therefore, as in Eq.~\eqref{eq_preconditoner}, the preconditioner $\hat{T}_l$ is introduced, and the factor $g_l$ is taken as
\begin{equation}
   g_l=0.4\frac{\lambda_l-(V+S)_{\rm min}}{(V+S)_{\rm min}}+0.6,
\end{equation}
with the single-particle energy $\lambda_l=\langle X_l|\hat{h}|X_l\rangle$.

\begin{figure}[!htbp]
  \centering
  \includegraphics[width=0.45\textwidth]{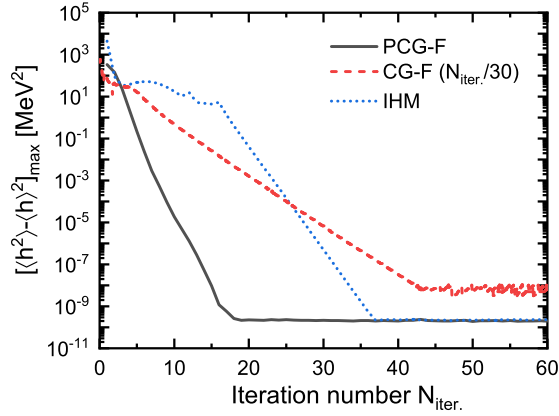}\\
  \caption{(Color online) Evolution of the maximum energy dispersion for the bound single-particle states in the spherical Woods-Saxon potential as a function of the iteration number.
  The solid and dashed lines respectively represent the results with and without the preconditioner, and the abscissa of the latter is scaled by a factor of $30$. The results of the inverse Hamiltonian method (dotted line) are also shown for comparison.
  }\label{fig2}
\end{figure}

To demonstrate the effects of the preconditioner, the PCG-F method is applied for a Dirac equation with a spherical Woods-Saxon potential in 3D lattice space by either considering the preconditioner $\hat{T}_l$ as in Eq.~\eqref{eq_preconditoner} or setting $\hat{T}_l$ as a unit operator, i.e., without preconditoner.
The evolution of the maximum energy dispersion $\langle \hat{h}^2\rangle-\langle \hat{h}\rangle^2$ for the bound single-particle states is shown in Fig.~\ref{fig2}.
For the PCG-F method, it takes only 15 iterations to reduce the maximum energy dispersion to $10^{-8}~{\rm MeV^2}$, while without the preconditioner, it takes more than 1200 iterations to reach the same level.
As a comparison, the IHM calculations require more than 30 iterations to reach this accuracy.
Moreover, one can see that the energy dispersions can drop to around $10^{-10}~{\rm MeV^2}$ after 18 iterations for the PCG-F method (after 37 iterations for the IHM), but they are finally fluctuated around $10^{-8}~{\rm MeV^2}$ for the calculations without the preconditioner.
Therefore, one can conclude that the preconditioner greatly improves the convergent accuracy and the speed of the iteration.

In the previous work with the IHM~\cite{REN2017Dirac3D}, the convergence of a wavefunction is regarded to be reached if the corresponding energy dispersion is smaller than $10^{-8}~{\rm MeV}^2$.
The same criterion is adopted in the present work, while $W_l^{(i)}$ or $P_l^{(i)}$ are removed from the summation in Eq.~\eqref{eq_Xip1} if the corresponding energy dispersion $\langle X_l | \hat{h}^2| X_l \rangle-\langle X_l |\hat{h} |X_l\rangle^2$ is smaller than $10^{-11}~{\rm MeV}^{2}$ or $10^{-8}~{\rm MeV}^{2}$, respectively.

\section{Results and discussion}
\subsection{Spherical potential}
We first assume the potentials $S$ and $V$ in Eq.~\eqref{Eq_potential} are spherical, and solve the corresponding Dirac equation with the PCG-F method.
The results of other methods, including the IHM in 3D lattice space and the shooting method in radial coordinate space, are used for comparison.
The numerical details used in the IHM is the same as those in the PCG-F method.
For the shooting method, the radial box size $R=20~{\rm fm}$ and the mesh size $dr=0.01~{\rm fm}$ are adopted, and the obtained results can be regarded as exact solutions thanks to the high accuracy.

\begin{figure}[!htbp]
  \centering
  \includegraphics[width=0.45\textwidth]{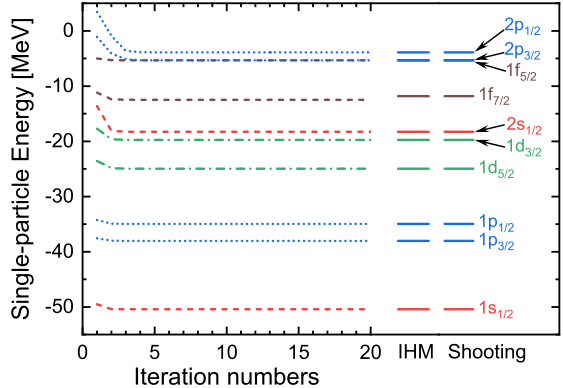}\\
  \caption{(Color online) Evolution of the single-particle energies in a spherical Woods-Saxon potential obtained by the PCG-F method as a function of the iteration numbers. For comparison, the single-particle energies obtained by the inverse Hamiltonian and the shooting methods are shown in the right side together with the spherical quantum numbers.
  }\label{fig3}
\end{figure}

In Fig.~\ref{fig3}, the evolution of single-particle energies obtained by the PCG-F method is shown as a function of the iteration numbers.
There are 40 bound states obtained in the PCG-F method, and they are grouped in energy according to the degeneracy due to the spherical symmetry.
One can see that the single-particle energies given by the PCG-F method are in very good agreement with those give by the IHM and the  shooting method.
Although the evolution in Fig.~\ref{fig3} is shown up to 20 iterations, to check the numerical stability of the PCG-F methods, the calculation was carried out up to the 3000th iteration.
It is found that the obtained single-particle energies of the bound states are quite stable after the 15th iteration, and the negative-energy states are eliminated perfectly.

\begin{figure}[!htbp]
  \centering
  \includegraphics[width=0.45\textwidth]{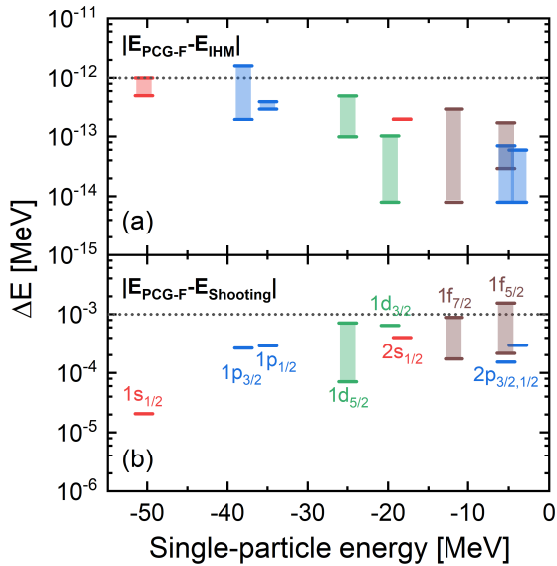}\\
  \caption{(Color online) Absolute differences between the single-particle energies in a spherical Woods-Saxon potential obtained with the PCG-F method and those with other methods including the inverse Hamiltonian (a) and shooting (b) methods.
  The spherical quantum numbers are listed in panel (b). For the degenerate states, only the maximum and minimum deviations are shown with the connecting bands.
  }\label{fig4}
\end{figure}

For a more precise comparison, Fig.~\ref{fig4} shows the absolute differences between the single-particle energies obtained with the PCG-F method and those with the inverse Hamiltonian and shooting methods.
In Fig.~\ref{fig4}(a), one can see that the absolute energy deviations between the PCG-F method and IHM are extremely small  for all states, i.e.,  less than $10^{-12}~{\rm MeV}$.
This demonstrates that the 3D lattice calculations with these two methods are in accuracy at the same level.

In Fig.~\ref{fig4}(b), the absolute energy deviations between the PCG-F and shooting methods are found to be in the range of $10^{-5}\sim10^{-3}~{\rm MeV}$.
This also shows the high accuracy of the PCG-F method, and it can be further improved by reducing the mesh size and/or enlarging the 3D box size.
Moreover, in contrast to the shooting solutions, the spherical degeneracy of the single-particle levels are slightly broken in the 3D lattice calculations.
This is because the discretized 3D lattice space is not exactly spherical, while the spherical symmetry is exactly fulfilled for the shooting method on radial coordinates.

\begin{figure}[!htbp]
  \centering
  \includegraphics[width=0.45\textwidth]{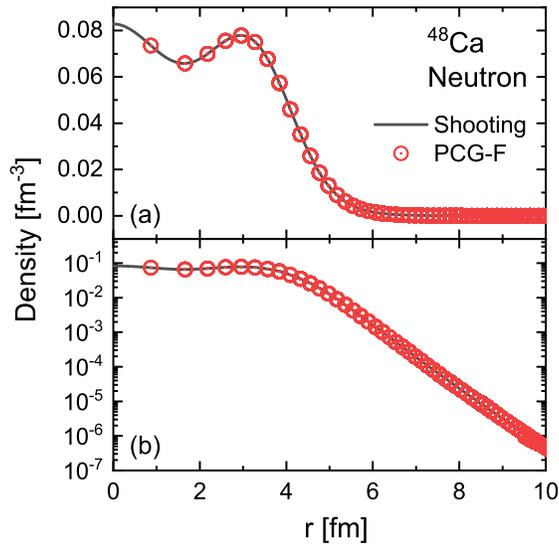}\\
  \caption{(Color online) The total density of the lowest 28 levels in the spherical Woods-Saxon potential as a function of the radial coordinate $r$ in normal (top) and logarithmic scales (bottom).
  The circles and lines represent the results of the PCG-F and shooting methods, respectively.
  }\label{fig5}
\end{figure}

Apart from the single-particle energies, it is necessary to examine the accuracy of the wavefunctions.
This is shown in Fig.~\ref{fig5}, where the total density of the lowest 28 levels, i.e., the neutron density for $^{48}$Ca, are shown as a function of the radial coordinate $r$ in normal and logarithmic scales, respectively.
The density obtained with the PCG-F method agrees with that with the shooting method very well, even at very large $r$ values.
In the region of $r>5$~fm, both densities decrease exponentially. This reveals that the PCG-F method could properly describe the asymptotic behavior of the single-particle wavefunctions.

\subsection{Deformed potential}
Deformation can be introduced to the potentials $S$ and $V$ through the parameters $(\beta,\gamma,\beta_{30})$ in Eq.~\eqref{Eq_deformparameter}.
We investigate three cases with the PCG-F method, i.e., $(\beta,\gamma,\beta_{30})=(0.3,0^\circ,0)$, $(0.3,30^\circ,0)$, and $(0.3,30^\circ,0.1)$, and they correspond to the axial, triaxial, and triaxially octupole potentials, respectively.
The calculated results are compared with the IHM in 3D lattice space.

\begin{figure}[!htbp]
  \centering
  \includegraphics[width=0.45\textwidth]{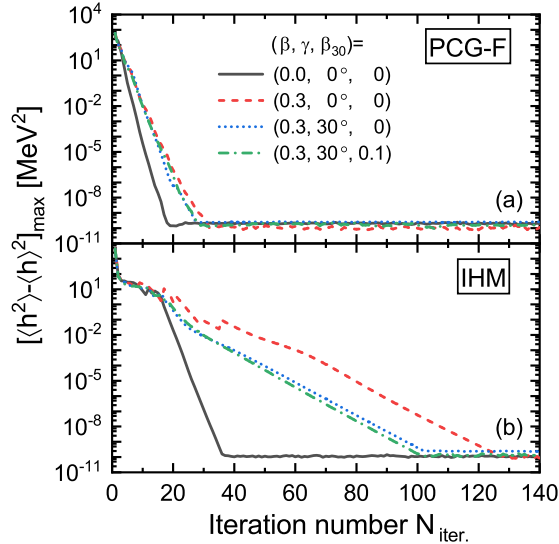}\\
  \caption{(Color online) The maximum energy dispersions of the bound single-particle states in various Woods-Saxon potentials obtained by the PCG-F (top) and inverse Hamiltonian methods (bottom) as a function of the iteration number.
  }\label{fig6}
\end{figure}

Firstly, the convergent behavior of the PCG-F and inverse Hamiltonian methods is examined for spherical and deformed potentials in Fig.~\ref{fig6}, where the maximum energy dispersions of the bound single-particle states are shown as a function of the iteration number.
For deformed potentials, the convergence for both the PCG-F and inverse Hamiltonian methods becomes slower.
This is because, on the one hand, there are more bound states in deformed potentials than in spherical ones, and, on the other hand, the initial guess of the wavefunctions is usually spherical for simplicity.

Nevertheless, for all potentials, the iteration of the PCG-F method is more efficient than the inverse Hamiltonian method.
In particular for deformed potentials, for instance, it takes less than 30 iterations for the PCG-F method to reduce the maximum energy dispersion to around $10^{-10}~{\rm MeV^2}$, while for the inverse Hamiltonian method, one needs more than 100 iterations.
This feature should be helpful to save the computational time for the future studies on many phenomena with the 3D lattice CDFT, such as exotic deformations~\cite{Zhao2015Rod-shaped, Zhao2018Spectroscopies, Zhao2017Phys.Rev.C14320}, super-heavy nuclei~\cite{Agbemava2015superheavy, Shi2019Superheavy, meng2020Hs270}, fission~\cite{Lu2012Potential, Lu2014MCRMF, Zhou2016PS}, 
fusion dynamics~\cite{Umar2015SHE_TDHF,Guo2018TDHF_fusion,Ren2020HeBeTDCDFT}, etc.

\begin{figure}[!htbp]
  \centering
  \includegraphics[width=0.45\textwidth]{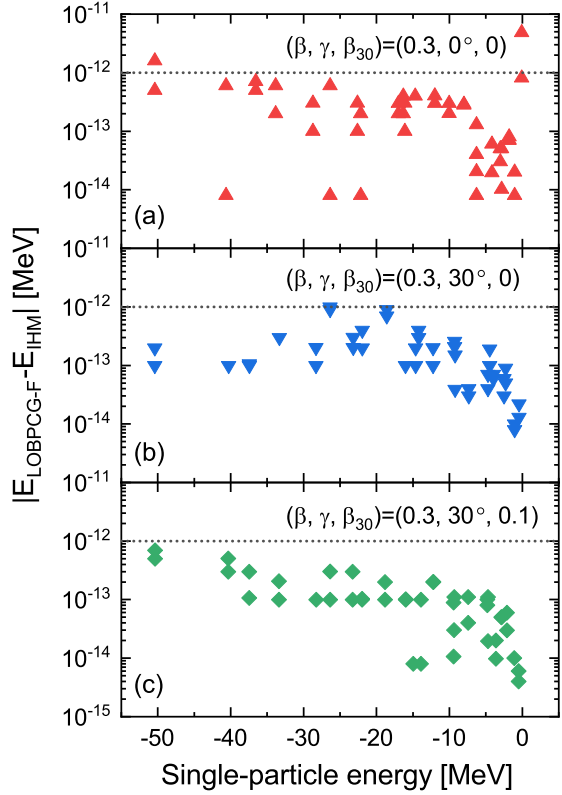}\\
  \caption{(Color online) Absolute differences between the single-particle energies in deformed Woods-Saxon potentials obtained with the PCG-F method and those with the inverse Hamiltonian method. Panels (a), (b), and (c) present the results in axial, triaxial, and triaxially octupole potentials, respectively.}\label{fig7}
\end{figure}

In Fig.~\ref{fig7}, the absolute differences between the single-particle energies obtained with the PCG-F method and those with the inverse Hamiltonian method are depicted.
Similar to the spherical case [see Fig.~\ref{fig4}(a)], the results of the two methods agree with each other in a very high accuracy, i.e., less than $10^{-12}~{\rm MeV}$.
In particular, the magnitudes of the deviations are not affected by the shape of the potential.
This demonstrates that the 3D lattice calculations realized by the PCG-F and the inverse Hamiltonian methods are essentially equivalent.
However, considering the high efficiency of the PCG-F method, to develop the CDFT in 3D lattice space with this method would be very beneficial in the future. 

Moreover, it is worthwhile to mention the perspective of the present PCG-F method on solving the RHB equation, where pairing correlations are taken into account with the Bogoliubov transformation.
Instead of diagonalizing huge matrices in the basis expansion method, the nonrelativistic HFB equation can be solved by the powerful gradient method, which is quite robust and easily deals with multiple constraints~\cite{ring2004nuclear}.
However, a direct application of the gradient method for the RHB equation is inhibited by the quasiparticle states in Dirac sea.
In this sense, the present PCG-F method, with an appropriate filtering function, seems a promising way to solve the relativistic Hartree-Bogoliubov equation iteratively.

\section{Summary}
In summary, an efficient method, PCG-F, has been proposed for solving nuclear Dirac equation in 3D lattice space, where a filtering function is adopted to avoid the variational collapsed problem and a momentum-dependent preconditioner is introduced to promote the efficiency of the iteration.
The method has been demonstrated in solving the Dirac equation with spherical and deformed Woods-Saxon potentials.
In the spherical case, the PCG-F method reproduces the single-particle energies and densities obtained by the shooting method in radial coordinate space with a high accuracy.
In both spherical and deformed cases, the single-particle energies obtained with the PCG-F and inverse Hamiltonian methods agree with each other very precisely, but the PCG-F method is much faster to achieve the convergence of the iteration, in particular for deformed potentials.
Considering the high efficiency of the PCG-F method, to develop the CDFT in 3D lattice space with this method would be very beneficial in the future. Moreover, the present PCG-F method seems a promising way to solve the relativistic Hartree-Bogoliubov equation iteratively.
Works following these directions are in progress.

\begin{acknowledgments}
This work was partly supported by the National Key R\&D Program of China (Contracts No. 2018YFA0404400 and 2017YFE0116700), the National Natural Science Foundation of China (Grants No. 11621131001, 11875075, 11935003, and 11975031), the State Key Laboratory of Nuclear Physics and Technology, Peking University (No. NPT2020ZZ01), and the China Postdoctoral Science Foundation under Grant No. 2020M670013.
\end{acknowledgments}

%\bibliography{reference}

\begin{thebibliography}{60}%
\makeatletter
\providecommand \@ifxundefined [1]{%
 \@ifx{#1\undefined}
}%
\providecommand \@ifnum [1]{%
 \ifnum #1\expandafter \@firstoftwo
 \else \expandafter \@secondoftwo
 \fi
}%
\providecommand \@ifx [1]{%
 \ifx #1\expandafter \@firstoftwo
 \else \expandafter \@secondoftwo
 \fi
}%
\providecommand \natexlab [1]{#1}%
\providecommand \enquote  [1]{``#1''}%
\providecommand \bibnamefont  [1]{#1}%
\providecommand \bibfnamefont [1]{#1}%
\providecommand \citenamefont [1]{#1}%
\providecommand \href@noop [0]{\@secondoftwo}%
\providecommand \href [0]{\begingroup \@sanitize@url \@href}%
\providecommand \@href[1]{\@@startlink{#1}\@@href}%
\providecommand \@@href[1]{\endgroup#1\@@endlink}%
\providecommand \@sanitize@url [0]{\catcode `\\12\catcode `\$12\catcode
  `\&12\catcode `\#12\catcode `\^12\catcode `\_12\catcode `\%12\relax}%
\providecommand \@@startlink[1]{}%
\providecommand \@@endlink[0]{}%
\providecommand \url  [0]{\begingroup\@sanitize@url \@url }%
\providecommand \@url [1]{\endgroup\@href {#1}{\urlprefix }}%
\providecommand \urlprefix  [0]{URL }%
\providecommand \Eprint [0]{\href }%
\providecommand \doibase [0]{http://dx.doi.org/}%
\providecommand \selectlanguage [0]{\@gobble}%
\providecommand \bibinfo  [0]{\@secondoftwo}%
\providecommand \bibfield  [0]{\@secondoftwo}%
\providecommand \translation [1]{[#1]}%
\providecommand \BibitemOpen [0]{}%
\providecommand \bibitemStop [0]{}%
\providecommand \bibitemNoStop [0]{.\EOS\space}%
\providecommand \EOS [0]{\spacefactor3000\relax}%
\providecommand \BibitemShut  [1]{\csname bibitem#1\endcsname}%
\let\auto@bib@innerbib\@empty
%</preamble>
\bibitem [{\citenamefont {Bender}\ \emph {et~al.}(2003)\citenamefont {Bender},
  \citenamefont {Heenen},\ and\ \citenamefont {Reinhard}}]{Bender2003Self}%
  \BibitemOpen
  \bibfield  {author} {\bibinfo {author} {\bibfnamefont {M.}~\bibnamefont
  {Bender}}, \bibinfo {author} {\bibfnamefont {P.-H.}\ \bibnamefont {Heenen}},
  \ and\ \bibinfo {author} {\bibfnamefont {P.-G.}\ \bibnamefont {Reinhard}},\
  }\href {\doibase 10.1103/RevModPhys.75.121} {\bibfield  {journal} {\bibinfo
  {journal} {Rev. Mod. Phys.}\ }\textbf {\bibinfo {volume} {75}},\ \bibinfo
  {pages} {121} (\bibinfo {year} {2003})}\BibitemShut {NoStop}%
\bibitem [{\citenamefont {Meng}(2016)}]{meng2016relativistic}%
  \BibitemOpen
  \bibinfo {editor} {\bibfnamefont {J.}~\bibnamefont {Meng}},\ ed.,\ \href@noop
  {} {\emph {\bibinfo {title} {Relativistic Density Functional for Nuclear
  Structure}}},\ \bibinfo {series} {International Review of Nuclear Physics},
  Vol.~\bibinfo {volume} {10}\ (\bibinfo  {publisher} {World Scientific,
  Singapore},\ \bibinfo {year} {2016})\BibitemShut {NoStop}%
\bibitem [{\citenamefont {Ring}(1996)}]{RING1996PPNP}%
  \BibitemOpen
  \bibfield  {author} {\bibinfo {author} {\bibfnamefont {P.}~\bibnamefont
  {Ring}},\ }\href {\doibase http://dx.doi.org/10.1016/0146-6410(96)00054-3}
  {\bibfield  {journal} {\bibinfo  {journal} {Prog. Part. Nucl. Phys.}\
  }\textbf {\bibinfo {volume} {37}},\ \bibinfo {pages} {193} (\bibinfo {year}
  {1996})}\BibitemShut {NoStop}%
\bibitem [{\citenamefont {Meng}\ \emph {et~al.}(2006)\citenamefont {Meng},
  \citenamefont {Toki}, \citenamefont {Zhou}, \citenamefont {Zhang},
  \citenamefont {Long},\ and\ \citenamefont {Geng}}]{meng2006PPNP}%
  \BibitemOpen
  \bibfield  {author} {\bibinfo {author} {\bibfnamefont {J.}~\bibnamefont
  {Meng}}, \bibinfo {author} {\bibfnamefont {H.}~\bibnamefont {Toki}}, \bibinfo
  {author} {\bibfnamefont {S.~G.}\ \bibnamefont {Zhou}}, \bibinfo {author}
  {\bibfnamefont {S.~Q.}\ \bibnamefont {Zhang}}, \bibinfo {author}
  {\bibfnamefont {W.~H.}\ \bibnamefont {Long}}, \ and\ \bibinfo {author}
  {\bibfnamefont {L.~S.}\ \bibnamefont {Geng}},\ }\href {\doibase
  https://doi.org/10.1016/j.ppnp.2005.06.001} {\bibfield  {journal} {\bibinfo
  {journal} {Prog. Part. Nucl. Phys.}\ }\textbf {\bibinfo {volume} {57}},\
  \bibinfo {pages} {470} (\bibinfo {year} {2006})}\BibitemShut {NoStop}%
\bibitem [{\citenamefont {Vretenar}\ \emph {et~al.}(2005)\citenamefont
  {Vretenar}, \citenamefont {Afanasjev}, \citenamefont {Lalazissis},\ and\
  \citenamefont {Ring}}]{Vretenar2005PhysicsReport}%
  \BibitemOpen
  \bibfield  {author} {\bibinfo {author} {\bibfnamefont {D.}~\bibnamefont
  {Vretenar}}, \bibinfo {author} {\bibfnamefont {A.~V.}\ \bibnamefont
  {Afanasjev}}, \bibinfo {author} {\bibfnamefont {G.~A.}\ \bibnamefont
  {Lalazissis}}, \ and\ \bibinfo {author} {\bibfnamefont {P.}~\bibnamefont
  {Ring}},\ }\href {\doibase http://dx.doi.org/10.1016/j.physrep.2004.10.001}
  {\bibfield  {journal} {\bibinfo  {journal} {Phys. Rep.}\ }\textbf {\bibinfo
  {volume} {409}},\ \bibinfo {pages} {101} (\bibinfo {year}
  {2005})}\BibitemShut {NoStop}%
\bibitem [{\citenamefont {Nik\ifmmode \check{s}\else
  \v{s}\fi{}i\ifmmode~\acute{c}\else \'{c}\fi{}}\ \emph
  {et~al.}(2011)\citenamefont {Nik\ifmmode \check{s}\else
  \v{s}\fi{}i\ifmmode~\acute{c}\else \'{c}\fi{}}, \citenamefont {Vretenar},\
  and\ \citenamefont {Ring}}]{NIKSIC2011PPNP}%
  \BibitemOpen
  \bibfield  {author} {\bibinfo {author} {\bibfnamefont {T.}~\bibnamefont
  {Nik\ifmmode \check{s}\else \v{s}\fi{}i\ifmmode~\acute{c}\else \'{c}\fi{}}},
  \bibinfo {author} {\bibfnamefont {D.}~\bibnamefont {Vretenar}}, \ and\
  \bibinfo {author} {\bibfnamefont {P.}~\bibnamefont {Ring}},\ }\href {\doibase
  https://doi.org/10.1016/j.ppnp.2011.01.055} {\bibfield  {journal} {\bibinfo
  {journal} {Prog. Part. Nucl. Phys.}\ }\textbf {\bibinfo {volume} {66}},\
  \bibinfo {pages} {519} (\bibinfo {year} {2011})}\BibitemShut {NoStop}%
\bibitem [{\citenamefont {Long}\ \emph {et~al.}(2004)\citenamefont {Long},
  \citenamefont {Meng}, \citenamefont {Van~Giai},\ and\ \citenamefont
  {Zhou}}]{Long2004PK1}%
  \BibitemOpen
  \bibfield  {author} {\bibinfo {author} {\bibfnamefont {W.}~\bibnamefont
  {Long}}, \bibinfo {author} {\bibfnamefont {J.}~\bibnamefont {Meng}}, \bibinfo
  {author} {\bibfnamefont {N.}~\bibnamefont {Van~Giai}}, \ and\ \bibinfo
  {author} {\bibfnamefont {S.-G.}\ \bibnamefont {Zhou}},\ }\href {\doibase
  10.1103/PhysRevC.69.034319} {\bibfield  {journal} {\bibinfo  {journal} {Phys.
  Rev. C}\ }\textbf {\bibinfo {volume} {69}},\ \bibinfo {pages} {034319}
  (\bibinfo {year} {2004})}\BibitemShut {NoStop}%
\bibitem [{\citenamefont {Lalazissis}\ \emph {et~al.}(2005)\citenamefont
  {Lalazissis}, \citenamefont {Nik\ifmmode \check{s}\else
  \v{s}\fi{}i\ifmmode~\acute{c}\else \'{c}\fi{}}, \citenamefont {Vretenar},\
  and\ \citenamefont {Ring}}]{lalazissis2005new}%
  \BibitemOpen
  \bibfield  {author} {\bibinfo {author} {\bibfnamefont {G.~A.}\ \bibnamefont
  {Lalazissis}}, \bibinfo {author} {\bibfnamefont {T.}~\bibnamefont
  {Nik\ifmmode \check{s}\else \v{s}\fi{}i\ifmmode~\acute{c}\else \'{c}\fi{}}},
  \bibinfo {author} {\bibfnamefont {D.}~\bibnamefont {Vretenar}}, \ and\
  \bibinfo {author} {\bibfnamefont {P.}~\bibnamefont {Ring}},\ }\href {\doibase
  10.1103/PhysRevC.71.024312} {\bibfield  {journal} {\bibinfo  {journal} {Phys.
  Rev. C}\ }\textbf {\bibinfo {volume} {71}},\ \bibinfo {pages} {024312}
  (\bibinfo {year} {2005})}\BibitemShut {NoStop}%
\bibitem [{\citenamefont {Nik\ifmmode \check{s}\else
  \v{s}\fi{}i\ifmmode~\acute{c}\else \'{c}\fi{}}\ \emph
  {et~al.}(2008)\citenamefont {Nik\ifmmode \check{s}\else
  \v{s}\fi{}i\ifmmode~\acute{c}\else \'{c}\fi{}}, \citenamefont {Vretenar},\
  and\ \citenamefont {Ring}}]{Niksic2008DDPC1}%
  \BibitemOpen
  \bibfield  {author} {\bibinfo {author} {\bibfnamefont {T.}~\bibnamefont
  {Nik\ifmmode \check{s}\else \v{s}\fi{}i\ifmmode~\acute{c}\else \'{c}\fi{}}},
  \bibinfo {author} {\bibfnamefont {D.}~\bibnamefont {Vretenar}}, \ and\
  \bibinfo {author} {\bibfnamefont {P.}~\bibnamefont {Ring}},\ }\href {\doibase
  10.1103/PhysRevC.78.034318} {\bibfield  {journal} {\bibinfo  {journal} {Phys.
  Rev. C}\ }\textbf {\bibinfo {volume} {78}},\ \bibinfo {pages} {034318}
  (\bibinfo {year} {2008})}\BibitemShut {NoStop}%
\bibitem [{\citenamefont {Zhao}\ \emph {et~al.}(2010)\citenamefont {Zhao},
  \citenamefont {Li}, \citenamefont {Yao},\ and\ \citenamefont
  {Meng}}]{ZhaoPC-PK1}%
  \BibitemOpen
  \bibfield  {author} {\bibinfo {author} {\bibfnamefont {P.~W.}\ \bibnamefont
  {Zhao}}, \bibinfo {author} {\bibfnamefont {Z.~P.}\ \bibnamefont {Li}},
  \bibinfo {author} {\bibfnamefont {J.~M.}\ \bibnamefont {Yao}}, \ and\
  \bibinfo {author} {\bibfnamefont {J.}~\bibnamefont {Meng}},\ }\href {\doibase
  10.1103/PhysRevC.82.054319} {\bibfield  {journal} {\bibinfo  {journal} {Phys.
  Rev. C}\ }\textbf {\bibinfo {volume} {82}},\ \bibinfo {pages} {054319}
  (\bibinfo {year} {2010})}\BibitemShut {NoStop}%
\bibitem [{\citenamefont {Sharma}\ \emph {et~al.}(1993)\citenamefont {Sharma},
  \citenamefont {Lalazissis},\ and\ \citenamefont
  {Ring}}]{Sharma1993Phys.Lett.B9}%
  \BibitemOpen
  \bibfield  {author} {\bibinfo {author} {\bibfnamefont {M.~M.}\ \bibnamefont
  {Sharma}}, \bibinfo {author} {\bibfnamefont {G.~A.}\ \bibnamefont
  {Lalazissis}}, \ and\ \bibinfo {author} {\bibfnamefont {P.}~\bibnamefont
  {Ring}},\ }\href {\doibase 10.1016/0370-2693(93)91561-Z} {\bibfield
  {journal} {\bibinfo  {journal} {Phys. Lett. B}\ }\textbf {\bibinfo {volume}
  {317}},\ \bibinfo {pages} {9} (\bibinfo {year} {1993})}\BibitemShut {NoStop}%
\bibitem [{\citenamefont {Meng}\ \emph {et~al.}(2013)\citenamefont {Meng},
  \citenamefont {Peng}, \citenamefont {Zhang},\ and\ \citenamefont
  {Zhao}}]{Meng2013FT_TAC}%
  \BibitemOpen
  \bibfield  {author} {\bibinfo {author} {\bibfnamefont {J.}~\bibnamefont
  {Meng}}, \bibinfo {author} {\bibfnamefont {J.}~\bibnamefont {Peng}}, \bibinfo
  {author} {\bibfnamefont {S.-Q.}\ \bibnamefont {Zhang}}, \ and\ \bibinfo
  {author} {\bibfnamefont {P.-W.}\ \bibnamefont {Zhao}},\ }\href {\doibase
  10.1007/s11467-013-0287-y} {\bibfield  {journal} {\bibinfo  {journal} {Front.
  Phys.}\ }\textbf {\bibinfo {volume} {8}},\ \bibinfo {pages} {55} (\bibinfo
  {year} {2013})}\BibitemShut {NoStop}%
\bibitem [{\citenamefont {Zhao}\ and\ \citenamefont
  {Li}(2018)}]{Zhao2018Spectroscopies}%
  \BibitemOpen
  \bibfield  {author} {\bibinfo {author} {\bibfnamefont {P.}~\bibnamefont
  {Zhao}}\ and\ \bibinfo {author} {\bibfnamefont {Z.}~\bibnamefont {Li}},\
  }\href {https://doi.org/10.1142/S0218301318300072} {\bibfield  {journal}
  {\bibinfo  {journal} {Int. J. Mod. Phys. E}\ }\textbf {\bibinfo {volume}
  {27}},\ \bibinfo {pages} {1830007} (\bibinfo {year} {2018})}\BibitemShut
  {NoStop}%
\bibitem [{\citenamefont {Tanihata}\ \emph {et~al.}(1985)\citenamefont
  {Tanihata}, \citenamefont {Hamagaki}, \citenamefont {Hashimoto},
  \citenamefont {Shida}, \citenamefont {Yoshikawa}, \citenamefont {Sugimoto},
  \citenamefont {Yamakawa}, \citenamefont {Kobayashi},\ and\ \citenamefont
  {Takahashi}}]{Tanihata1985halo}%
  \BibitemOpen
  \bibfield  {author} {\bibinfo {author} {\bibfnamefont {I.}~\bibnamefont
  {Tanihata}}, \bibinfo {author} {\bibfnamefont {H.}~\bibnamefont {Hamagaki}},
  \bibinfo {author} {\bibfnamefont {O.}~\bibnamefont {Hashimoto}}, \bibinfo
  {author} {\bibfnamefont {Y.}~\bibnamefont {Shida}}, \bibinfo {author}
  {\bibfnamefont {N.}~\bibnamefont {Yoshikawa}}, \bibinfo {author}
  {\bibfnamefont {K.}~\bibnamefont {Sugimoto}}, \bibinfo {author}
  {\bibfnamefont {O.}~\bibnamefont {Yamakawa}}, \bibinfo {author}
  {\bibfnamefont {T.}~\bibnamefont {Kobayashi}}, \ and\ \bibinfo {author}
  {\bibfnamefont {N.}~\bibnamefont {Takahashi}},\ }\href {\doibase
  10.1103/PhysRevLett.55.2676} {\bibfield  {journal} {\bibinfo  {journal}
  {Phys. Rev. Lett.}\ }\textbf {\bibinfo {volume} {55}},\ \bibinfo {pages}
  {2676} (\bibinfo {year} {1985})}\BibitemShut {NoStop}%
\bibitem [{\citenamefont {Meng}\ and\ \citenamefont
  {Zhou}(2015)}]{meng2015halos}%
  \BibitemOpen
  \bibfield  {author} {\bibinfo {author} {\bibfnamefont {J.}~\bibnamefont
  {Meng}}\ and\ \bibinfo {author} {\bibfnamefont {S.-G.}\ \bibnamefont
  {Zhou}},\ }\href@noop {} {\bibfield  {journal} {\bibinfo  {journal} {J. Phys.
  G}\ }\textbf {\bibinfo {volume} {42}},\ \bibinfo {pages} {093101} (\bibinfo
  {year} {2015})}\BibitemShut {NoStop}%
\bibitem [{\citenamefont {Meng}(1998)}]{meng1998NPA}%
  \BibitemOpen
  \bibfield  {author} {\bibinfo {author} {\bibfnamefont {J.}~\bibnamefont
  {Meng}},\ }\href {\doibase https://doi.org/10.1016/S0375-9474(98)00178-X}
  {\bibfield  {journal} {\bibinfo  {journal} {Nucl. Phys. A}\ }\textbf
  {\bibinfo {volume} {635}},\ \bibinfo {pages} {3 } (\bibinfo {year}
  {1998})}\BibitemShut {NoStop}%
\bibitem [{\citenamefont {Meng}\ and\ \citenamefont
  {Ring}(1996)}]{meng1996relativistic}%
  \BibitemOpen
  \bibfield  {author} {\bibinfo {author} {\bibfnamefont {J.}~\bibnamefont
  {Meng}}\ and\ \bibinfo {author} {\bibfnamefont {P.}~\bibnamefont {Ring}},\
  }\href {\doibase 10.1103/PhysRevLett.77.3963} {\bibfield  {journal} {\bibinfo
   {journal} {Phys. Rev. Lett.}\ }\textbf {\bibinfo {volume} {77}},\ \bibinfo
  {pages} {3963} (\bibinfo {year} {1996})}\BibitemShut {NoStop}%
\bibitem [{\citenamefont {Meng}\ and\ \citenamefont
  {Ring}(1998)}]{meng1998giant}%
  \BibitemOpen
  \bibfield  {author} {\bibinfo {author} {\bibfnamefont {J.}~\bibnamefont
  {Meng}}\ and\ \bibinfo {author} {\bibfnamefont {P.}~\bibnamefont {Ring}},\
  }\href {\doibase 10.1103/PhysRevLett.80.460} {\bibfield  {journal} {\bibinfo
  {journal} {Phys. Rev. Lett.}\ }\textbf {\bibinfo {volume} {80}},\ \bibinfo
  {pages} {460} (\bibinfo {year} {1998})}\BibitemShut {NoStop}%
\bibitem [{\citenamefont {Xia}\ \emph {et~al.}(2018)\citenamefont {Xia},
  \citenamefont {Lim}, \citenamefont {Zhao}, \citenamefont {Liang},
  \citenamefont {Qu}, \citenamefont {Chen}, \citenamefont {Liu}, \citenamefont
  {Zhang}, \citenamefont {Zhang}, \citenamefont {Kim},\ and\ \citenamefont
  {Meng}}]{xia2018ADNDT}%
  \BibitemOpen
  \bibfield  {author} {\bibinfo {author} {\bibfnamefont {X.~W.}\ \bibnamefont
  {Xia}}, \bibinfo {author} {\bibfnamefont {Y.}~\bibnamefont {Lim}}, \bibinfo
  {author} {\bibfnamefont {P.~W.}\ \bibnamefont {Zhao}}, \bibinfo {author}
  {\bibfnamefont {H.~Z.}\ \bibnamefont {Liang}}, \bibinfo {author}
  {\bibfnamefont {X.~Y.}\ \bibnamefont {Qu}}, \bibinfo {author} {\bibfnamefont
  {Y.}~\bibnamefont {Chen}}, \bibinfo {author} {\bibfnamefont {H.}~\bibnamefont
  {Liu}}, \bibinfo {author} {\bibfnamefont {L.~F.}\ \bibnamefont {Zhang}},
  \bibinfo {author} {\bibfnamefont {S.~Q.}\ \bibnamefont {Zhang}}, \bibinfo
  {author} {\bibfnamefont {Y.}~\bibnamefont {Kim}}, \ and\ \bibinfo {author}
  {\bibfnamefont {J.}~\bibnamefont {Meng}},\ }\href {\doibase
  https://doi.org/10.1016/j.adt.2017.09.001} {\bibfield  {journal} {\bibinfo
  {journal} {At. Data Nucl. Data Tables}\ }\textbf {\bibinfo {volume}
  {121-122}},\ \bibinfo {pages} {1} (\bibinfo {year} {2018})}\BibitemShut
  {NoStop}%
\bibitem [{\citenamefont {Price}\ and\ \citenamefont
  {Walker}(1987)}]{Price1987Self-consistent}%
  \BibitemOpen
  \bibfield  {author} {\bibinfo {author} {\bibfnamefont {C.~E.}\ \bibnamefont
  {Price}}\ and\ \bibinfo {author} {\bibfnamefont {G.~E.}\ \bibnamefont
  {Walker}},\ }\href {\doibase 10.1103/PhysRevC.36.354} {\bibfield  {journal}
  {\bibinfo  {journal} {Phys. Rev. C}\ }\textbf {\bibinfo {volume} {36}},\
  \bibinfo {pages} {354} (\bibinfo {year} {1987})}\BibitemShut {NoStop}%
\bibitem [{\citenamefont {Zhou}\ \emph {et~al.}(2003)\citenamefont {Zhou},
  \citenamefont {Meng},\ and\ \citenamefont {Ring}}]{zhou2003spherical}%
  \BibitemOpen
  \bibfield  {author} {\bibinfo {author} {\bibfnamefont {S.-G.}\ \bibnamefont
  {Zhou}}, \bibinfo {author} {\bibfnamefont {J.}~\bibnamefont {Meng}}, \ and\
  \bibinfo {author} {\bibfnamefont {P.}~\bibnamefont {Ring}},\ }\href {\doibase
  10.1103/PhysRevC.68.034323} {\bibfield  {journal} {\bibinfo  {journal} {Phys.
  Rev. C}\ }\textbf {\bibinfo {volume} {68}},\ \bibinfo {pages} {034323}
  (\bibinfo {year} {2003})}\BibitemShut {NoStop}%
\bibitem [{\citenamefont {Zhou}\ \emph {et~al.}(2010)\citenamefont {Zhou},
  \citenamefont {Meng}, \citenamefont {Ring},\ and\ \citenamefont
  {Zhao}}]{zhou2010neutron}%
  \BibitemOpen
  \bibfield  {author} {\bibinfo {author} {\bibfnamefont {S.-G.}\ \bibnamefont
  {Zhou}}, \bibinfo {author} {\bibfnamefont {J.}~\bibnamefont {Meng}}, \bibinfo
  {author} {\bibfnamefont {P.}~\bibnamefont {Ring}}, \ and\ \bibinfo {author}
  {\bibfnamefont {E.-G.}\ \bibnamefont {Zhao}},\ }\href {\doibase
  10.1103/PhysRevC.82.011301} {\bibfield  {journal} {\bibinfo  {journal} {Phys.
  Rev. C}\ }\textbf {\bibinfo {volume} {82}},\ \bibinfo {pages} {011301}
  (\bibinfo {year} {2010})}\BibitemShut {NoStop}%
\bibitem [{\citenamefont {Li}\ \emph {et~al.}(2012)\citenamefont {Li},
  \citenamefont {Meng}, \citenamefont {Ring}, \citenamefont {Zhao},\ and\
  \citenamefont {Zhou}}]{li2012deformed}%
  \BibitemOpen
  \bibfield  {author} {\bibinfo {author} {\bibfnamefont {L.}~\bibnamefont
  {Li}}, \bibinfo {author} {\bibfnamefont {J.}~\bibnamefont {Meng}}, \bibinfo
  {author} {\bibfnamefont {P.}~\bibnamefont {Ring}}, \bibinfo {author}
  {\bibfnamefont {E.-G.}\ \bibnamefont {Zhao}}, \ and\ \bibinfo {author}
  {\bibfnamefont {S.-G.}\ \bibnamefont {Zhou}},\ }\href {\doibase
  10.1103/PhysRevC.85.024312} {\bibfield  {journal} {\bibinfo  {journal} {Phys.
  Rev. C}\ }\textbf {\bibinfo {volume} {85}},\ \bibinfo {pages} {024312}
  (\bibinfo {year} {2012})}\BibitemShut {NoStop}%
\bibitem [{\citenamefont {Sun}\ \emph {et~al.}(2018)\citenamefont {Sun},
  \citenamefont {Zhao},\ and\ \citenamefont {Zhou}}]{SUN2018C22}%
  \BibitemOpen
  \bibfield  {author} {\bibinfo {author} {\bibfnamefont {X.-X.}\ \bibnamefont
  {Sun}}, \bibinfo {author} {\bibfnamefont {J.}~\bibnamefont {Zhao}}, \ and\
  \bibinfo {author} {\bibfnamefont {S.-G.}\ \bibnamefont {Zhou}},\ }\href
  {\doibase https://doi.org/10.1016/j.physletb.2018.08.071} {\bibfield
  {journal} {\bibinfo  {journal} {Phys. Lett. B}\ }\textbf {\bibinfo {volume}
  {785}},\ \bibinfo {pages} {530} (\bibinfo {year} {2018})}\BibitemShut
  {NoStop}%
\bibitem [{\citenamefont {Zhang}\ \emph {et~al.}(2019)\citenamefont {Zhang},
  \citenamefont {Wang},\ and\ \citenamefont {Zhang}}]{zhang2019forbidden}%
  \BibitemOpen
  \bibfield  {author} {\bibinfo {author} {\bibfnamefont {K.~Y.}\ \bibnamefont
  {Zhang}}, \bibinfo {author} {\bibfnamefont {D.~Y.}\ \bibnamefont {Wang}}, \
  and\ \bibinfo {author} {\bibfnamefont {S.~Q.}\ \bibnamefont {Zhang}},\ }\href
  {\doibase 10.1103/PhysRevC.100.034312} {\bibfield  {journal} {\bibinfo
  {journal} {Phys. Rev. C}\ }\textbf {\bibinfo {volume} {100}},\ \bibinfo
  {pages} {034312} (\bibinfo {year} {2019})}\BibitemShut {NoStop}%
\bibitem [{\citenamefont {Long}\ \emph
  {et~al.}(2010{\natexlab{a}})\citenamefont {Long}, \citenamefont {Ring},
  \citenamefont {Giai},\ and\ \citenamefont {Meng}}]{Long2010RHFB}%
  \BibitemOpen
  \bibfield  {author} {\bibinfo {author} {\bibfnamefont {W.~H.}\ \bibnamefont
  {Long}}, \bibinfo {author} {\bibfnamefont {P.}~\bibnamefont {Ring}}, \bibinfo
  {author} {\bibfnamefont {N.~V.}\ \bibnamefont {Giai}}, \ and\ \bibinfo
  {author} {\bibfnamefont {J.}~\bibnamefont {Meng}},\ }\href {\doibase
  10.1103/PhysRevC.81.024308} {\bibfield  {journal} {\bibinfo  {journal} {Phys.
  Rev. C}\ }\textbf {\bibinfo {volume} {81}},\ \bibinfo {pages} {024308}
  (\bibinfo {year} {2010}{\natexlab{a}})}\BibitemShut {NoStop}%
\bibitem [{\citenamefont {Long}\ \emph
  {et~al.}(2010{\natexlab{b}})\citenamefont {Long}, \citenamefont {Ring},
  \citenamefont {Meng}, \citenamefont {Van~Giai},\ and\ \citenamefont
  {Bertulani}}]{Long2010RHFB_halo_pss}%
  \BibitemOpen
  \bibfield  {author} {\bibinfo {author} {\bibfnamefont {W.~H.}\ \bibnamefont
  {Long}}, \bibinfo {author} {\bibfnamefont {P.}~\bibnamefont {Ring}}, \bibinfo
  {author} {\bibfnamefont {J.}~\bibnamefont {Meng}}, \bibinfo {author}
  {\bibfnamefont {N.}~\bibnamefont {Van~Giai}}, \ and\ \bibinfo {author}
  {\bibfnamefont {C.~A.}\ \bibnamefont {Bertulani}},\ }\href {\doibase
  10.1103/PhysRevC.81.031302} {\bibfield  {journal} {\bibinfo  {journal} {Phys.
  Rev. C}\ }\textbf {\bibinfo {volume} {81}},\ \bibinfo {pages} {031302(R)}
  (\bibinfo {year} {2010}{\natexlab{b}})}\BibitemShut {NoStop}%
\bibitem [{\citenamefont {Geng}\ \emph {et~al.}(2020)\citenamefont {Geng},
  \citenamefont {Xiang}, \citenamefont {Sun},\ and\ \citenamefont
  {Long}}]{geng2020defRHF}%
  \BibitemOpen
  \bibfield  {author} {\bibinfo {author} {\bibfnamefont {J.}~\bibnamefont
  {Geng}}, \bibinfo {author} {\bibfnamefont {J.}~\bibnamefont {Xiang}},
  \bibinfo {author} {\bibfnamefont {B.~Y.}\ \bibnamefont {Sun}}, \ and\
  \bibinfo {author} {\bibfnamefont {W.~H.}\ \bibnamefont {Long}},\ }\href
  {\doibase 10.1103/PhysRevC.101.064302} {\bibfield  {journal} {\bibinfo
  {journal} {Phys. Rev. C}\ }\textbf {\bibinfo {volume} {101}},\ \bibinfo
  {pages} {064302} (\bibinfo {year} {2020})}\BibitemShut {NoStop}%
\bibitem [{\citenamefont {Davies}\ \emph {et~al.}(1980)\citenamefont {Davies},
  \citenamefont {Flocard}, \citenamefont {Krieger},\ and\ \citenamefont
  {Weiss}}]{davies1980imaginary}%
  \BibitemOpen
  \bibfield  {author} {\bibinfo {author} {\bibfnamefont {K.~T.~R.}\
  \bibnamefont {Davies}}, \bibinfo {author} {\bibfnamefont {H.}~\bibnamefont
  {Flocard}}, \bibinfo {author} {\bibfnamefont {S.}~\bibnamefont {Krieger}}, \
  and\ \bibinfo {author} {\bibfnamefont {M.~S.}\ \bibnamefont {Weiss}},\ }\href
  {\doibase https://doi.org/10.1016/0375-9474(80)90509-6} {\bibfield  {journal}
  {\bibinfo  {journal} {Nucl. Phys. A}\ }\textbf {\bibinfo {volume} {342}},\
  \bibinfo {pages} {111} (\bibinfo {year} {1980})}\BibitemShut {NoStop}%
\bibitem [{\citenamefont {Bonche}\ \emph {et~al.}(2005)\citenamefont {Bonche},
  \citenamefont {Flocard},\ and\ \citenamefont {Heenen}}]{bonche2005solution}%
  \BibitemOpen
  \bibfield  {author} {\bibinfo {author} {\bibfnamefont {P.}~\bibnamefont
  {Bonche}}, \bibinfo {author} {\bibfnamefont {H.}~\bibnamefont {Flocard}}, \
  and\ \bibinfo {author} {\bibfnamefont {P.}~\bibnamefont {Heenen}},\ }\href
  {\doibase https://doi.org/10.1016/j.cpc.2005.05.001} {\bibfield  {journal}
  {\bibinfo  {journal} {Comput. Phys. Commun.}\ }\textbf {\bibinfo {volume}
  {171}},\ \bibinfo {pages} {49 } (\bibinfo {year} {2005})}\BibitemShut
  {NoStop}%
\bibitem [{\citenamefont {Reinhard}\ and\ \citenamefont
  {Cusson}(1982)}]{Reinhard1982comparative}%
  \BibitemOpen
  \bibfield  {author} {\bibinfo {author} {\bibfnamefont {P.-G.}\ \bibnamefont
  {Reinhard}}\ and\ \bibinfo {author} {\bibfnamefont {R.}~\bibnamefont
  {Cusson}},\ }\href {\doibase https://doi.org/10.1016/0375-9474(82)90458-4}
  {\bibfield  {journal} {\bibinfo  {journal} {Nucl. Phys. A}\ }\textbf
  {\bibinfo {volume} {378}},\ \bibinfo {pages} {418} (\bibinfo {year}
  {1982})}\BibitemShut {NoStop}%
\bibitem [{\citenamefont {Maruhn}\ \emph {et~al.}(2014)\citenamefont {Maruhn},
  \citenamefont {Reinhard}, \citenamefont {Stevenson},\ and\ \citenamefont
  {Umar}}]{Maruhn2014CPC}%
  \BibitemOpen
  \bibfield  {author} {\bibinfo {author} {\bibfnamefont {J.~A.}\ \bibnamefont
  {Maruhn}}, \bibinfo {author} {\bibfnamefont {P.~G.}\ \bibnamefont
  {Reinhard}}, \bibinfo {author} {\bibfnamefont {P.~D.}\ \bibnamefont
  {Stevenson}}, \ and\ \bibinfo {author} {\bibfnamefont {A.~S.}\ \bibnamefont
  {Umar}},\ }\href {\doibase http://dx.doi.org/10.1016/j.cpc.2014.04.008}
  {\bibfield  {journal} {\bibinfo  {journal} {Comput. Phys. Commun.}\ }\textbf
  {\bibinfo {volume} {185}},\ \bibinfo {pages} {2195} (\bibinfo {year}
  {2014})}\BibitemShut {NoStop}%
\bibitem [{\citenamefont {Wang}\ and\ \citenamefont
  {Ren}(2018)}]{Wang2018Nilsson3D}%
  \BibitemOpen
  \bibfield  {author} {\bibinfo {author} {\bibfnamefont {Y.~Y.}\ \bibnamefont
  {Wang}}\ and\ \bibinfo {author} {\bibfnamefont {Z.~X.}\ \bibnamefont {Ren}},\
  }\href {\doibase 10.1007/s11433-018-9213-7} {\bibfield  {journal} {\bibinfo
  {journal} {Sci. China-Phys. Mech. Astron.}\ }\textbf {\bibinfo {volume}
  {61}},\ \bibinfo {pages} {082012} (\bibinfo {year} {2018})}\BibitemShut
  {NoStop}%
\bibitem [{\citenamefont {Zhang}\ \emph {et~al.}(2009)\citenamefont {Zhang},
  \citenamefont {Liang},\ and\ \citenamefont {Meng}}]{zhang2009first}%
  \BibitemOpen
  \bibfield  {author} {\bibinfo {author} {\bibfnamefont {Y.}~\bibnamefont
  {Zhang}}, \bibinfo {author} {\bibfnamefont {H.~Z.}\ \bibnamefont {Liang}}, \
  and\ \bibinfo {author} {\bibfnamefont {J.}~\bibnamefont {Meng}},\ }\href
  {http://stacks.iop.org/1674-1137/33/i=S1/a=036} {\bibfield  {journal}
  {\bibinfo  {journal} {Chin. Phys. C}\ }\textbf {\bibinfo {volume} {33}},\
  \bibinfo {pages} {113} (\bibinfo {year} {2009})}\BibitemShut {NoStop}%
\bibitem [{\citenamefont {Zhang}\ \emph {et~al.}(2010)\citenamefont {Zhang},
  \citenamefont {Liang},\ and\ \citenamefont {Meng}}]{ZhangIJMPE2010}%
  \BibitemOpen
  \bibfield  {author} {\bibinfo {author} {\bibfnamefont {Y.}~\bibnamefont
  {Zhang}}, \bibinfo {author} {\bibfnamefont {H.}~\bibnamefont {Liang}}, \ and\
  \bibinfo {author} {\bibfnamefont {J.}~\bibnamefont {Meng}},\ }\href {\doibase
  10.1142/S0218301310014637} {\bibfield  {journal} {\bibinfo  {journal} {Int.
  J. Mod. Phys. E}\ }\textbf {\bibinfo {volume} {19}},\ \bibinfo {pages} {55}
  (\bibinfo {year} {2010})}\BibitemShut {NoStop}%
\bibitem [{\citenamefont {Hagino}\ and\ \citenamefont
  {Tanimura}(2010)}]{hagino2010iterative}%
  \BibitemOpen
  \bibfield  {author} {\bibinfo {author} {\bibfnamefont {K.}~\bibnamefont
  {Hagino}}\ and\ \bibinfo {author} {\bibfnamefont {Y.}~\bibnamefont
  {Tanimura}},\ }\href {\doibase 10.1103/PhysRevC.82.057301} {\bibfield
  {journal} {\bibinfo  {journal} {Phys. Rev. C}\ }\textbf {\bibinfo {volume}
  {82}},\ \bibinfo {pages} {057301} (\bibinfo {year} {2010})}\BibitemShut
  {NoStop}%
\bibitem [{\citenamefont {Tanimura}\ \emph {et~al.}(2015)\citenamefont
  {Tanimura}, \citenamefont {Hagino},\ and\ \citenamefont
  {Liang}}]{tanimura20153d}%
  \BibitemOpen
  \bibfield  {author} {\bibinfo {author} {\bibfnamefont {Y.}~\bibnamefont
  {Tanimura}}, \bibinfo {author} {\bibfnamefont {K.}~\bibnamefont {Hagino}}, \
  and\ \bibinfo {author} {\bibfnamefont {H.~Z.}\ \bibnamefont {Liang}},\ }\href
  {\doibase 10.1093/ptep/ptv083} {\bibfield  {journal} {\bibinfo  {journal}
  {Prog. Theor. Exp. Phys.}\ }\textbf {\bibinfo {volume} {2015}},\ \bibinfo
  {pages} {073D01} (\bibinfo {year} {2015})}\BibitemShut {NoStop}%
\bibitem [{\citenamefont {Ren}\ \emph {et~al.}(2017)\citenamefont {Ren},
  \citenamefont {Zhang},\ and\ \citenamefont {Meng}}]{REN2017Dirac3D}%
  \BibitemOpen
  \bibfield  {author} {\bibinfo {author} {\bibfnamefont {Z.~X.}\ \bibnamefont
  {Ren}}, \bibinfo {author} {\bibfnamefont {S.~Q.}\ \bibnamefont {Zhang}}, \
  and\ \bibinfo {author} {\bibfnamefont {J.}~\bibnamefont {Meng}},\ }\href
  {\doibase 10.1103/PhysRevC.95.024313} {\bibfield  {journal} {\bibinfo
  {journal} {Phys. Rev. C}\ }\textbf {\bibinfo {volume} {95}},\ \bibinfo
  {pages} {024313} (\bibinfo {year} {2017})}\BibitemShut {NoStop}%
\bibitem [{\citenamefont {Ren}\ \emph {et~al.}(2019)\citenamefont {Ren},
  \citenamefont {Zhang}, \citenamefont {Zhao}, \citenamefont {Itagaki},
  \citenamefont {Maruhn},\ and\ \citenamefont {Meng}}]{Ren2019C12LCS}%
  \BibitemOpen
  \bibfield  {author} {\bibinfo {author} {\bibfnamefont {Z.~X.}\ \bibnamefont
  {Ren}}, \bibinfo {author} {\bibfnamefont {S.~Q.}\ \bibnamefont {Zhang}},
  \bibinfo {author} {\bibfnamefont {P.~W.}\ \bibnamefont {Zhao}}, \bibinfo
  {author} {\bibfnamefont {N.}~\bibnamefont {Itagaki}}, \bibinfo {author}
  {\bibfnamefont {J.~A.}\ \bibnamefont {Maruhn}}, \ and\ \bibinfo {author}
  {\bibfnamefont {J.}~\bibnamefont {Meng}},\ }\href {\doibase
  https://doi.org/10.1007/s11433-019-9412-3} {\bibfield  {journal} {\bibinfo
  {journal} {Sci. China-Phys. Mech. Astron.}\ }\textbf {\bibinfo {volume}
  {62}},\ \bibinfo {pages} {112062} (\bibinfo {year} {2019})}\BibitemShut
  {NoStop}%
\bibitem [{\citenamefont {Ren}\ \emph {et~al.}(2020{\natexlab{a}})\citenamefont
  {Ren}, \citenamefont {Zhao}, \citenamefont {Zhang},\ and\ \citenamefont
  {Meng}}]{REN2020Toroidal}%
  \BibitemOpen
  \bibfield  {author} {\bibinfo {author} {\bibfnamefont {Z.~X.}\ \bibnamefont
  {Ren}}, \bibinfo {author} {\bibfnamefont {P.~W.}\ \bibnamefont {Zhao}},
  \bibinfo {author} {\bibfnamefont {S.~Q.}\ \bibnamefont {Zhang}}, \ and\
  \bibinfo {author} {\bibfnamefont {J.}~\bibnamefont {Meng}},\ }\href {\doibase
  https://doi.org/10.1016/j.nuclphysa.2020.121696} {\bibfield  {journal}
  {\bibinfo  {journal} {Nucl. Phys. A}\ }\textbf {\bibinfo {volume} {996}},\
  \bibinfo {pages} {121696} (\bibinfo {year} {2020}{\natexlab{a}})}\BibitemShut
  {NoStop}%
\bibitem [{\citenamefont {Ren}\ \emph {et~al.}(2020{\natexlab{b}})\citenamefont
  {Ren}, \citenamefont {Zhao},\ and\ \citenamefont {Meng}}]{Ren2020HeBeTDCDFT}%
  \BibitemOpen
  \bibfield  {author} {\bibinfo {author} {\bibfnamefont {Z.~X.}\ \bibnamefont
  {Ren}}, \bibinfo {author} {\bibfnamefont {P.~W.}\ \bibnamefont {Zhao}}, \
  and\ \bibinfo {author} {\bibfnamefont {J.}~\bibnamefont {Meng}},\ }\href
  {\doibase https://doi.org/10.1016/j.physletb.2019.135194} {\bibfield
  {journal} {\bibinfo  {journal} {Phys. Lett. B}\ }\textbf {\bibinfo {volume}
  {801}},\ \bibinfo {pages} {135194} (\bibinfo {year}
  {2020}{\natexlab{b}})}\BibitemShut {NoStop}%
\bibitem [{\citenamefont {Tanimura}(2014)}]{Tanimura2014PhdThesis}%
  \BibitemOpen
  \bibfield  {author} {\bibinfo {author} {\bibfnamefont {Y.}~\bibnamefont
  {Tanimura}},\ }\emph {\bibinfo {title} {Covariant Density Functional
  Calculations for Atomic Nuclei in the 3-dimensional Coordinate Space}},\
  \href@noop {} {Ph.D. thesis},\ \bibinfo  {school} {Department of Physics,
  Tohoku University} (\bibinfo {year} {2014})\BibitemShut {NoStop}%
\bibitem [{\citenamefont {Tanimura}\ \emph {et~al.}(2013)\citenamefont
  {Tanimura}, \citenamefont {Hagino},\ and\ \citenamefont
  {Ring}}]{Tanimura2013HFB_IHM}%
  \BibitemOpen
  \bibfield  {author} {\bibinfo {author} {\bibfnamefont {Y.}~\bibnamefont
  {Tanimura}}, \bibinfo {author} {\bibfnamefont {K.}~\bibnamefont {Hagino}}, \
  and\ \bibinfo {author} {\bibfnamefont {P.}~\bibnamefont {Ring}},\ }\href
  {\doibase 10.1103/PhysRevC.88.017301} {\bibfield  {journal} {\bibinfo
  {journal} {Phys. Rev. C}\ }\textbf {\bibinfo {volume} {88}},\ \bibinfo
  {pages} {017301} (\bibinfo {year} {2013})}\BibitemShut {NoStop}%
\bibitem [{\citenamefont {Lin}\ \emph {et~al.}(2013)\citenamefont {Lin},
  \citenamefont {Shao},\ and\ \citenamefont {E}}]{LIN2013LOBPCG-F}%
  \BibitemOpen
  \bibfield  {author} {\bibinfo {author} {\bibfnamefont {L.}~\bibnamefont
  {Lin}}, \bibinfo {author} {\bibfnamefont {S.}~\bibnamefont {Shao}}, \ and\
  \bibinfo {author} {\bibfnamefont {W.}~\bibnamefont {E}},\ }\href {\doibase
  https://doi.org/10.1016/j.jcp.2013.03.030} {\bibfield  {journal} {\bibinfo
  {journal} {J. Comput. Phys.}\ }\textbf {\bibinfo {volume} {245}},\ \bibinfo
  {pages} {205 } (\bibinfo {year} {2013})}\BibitemShut {NoStop}%
\bibitem [{\citenamefont {Ring}\ and\ \citenamefont
  {Schuck}(2004)}]{ring2004nuclear}%
  \BibitemOpen
  \bibfield  {author} {\bibinfo {author} {\bibfnamefont {P.}~\bibnamefont
  {Ring}}\ and\ \bibinfo {author} {\bibfnamefont {P.}~\bibnamefont {Schuck}},\
  }\href@noop {} {\emph {\bibinfo {title} {The nuclear many-body problem}}}\
  (\bibinfo  {publisher} {Springer Science \& Business Media},\ \bibinfo {year}
  {2004})\BibitemShut {NoStop}%
\bibitem [{\citenamefont {Hestenes}\ and\ \citenamefont
  {Stiefel}(1952)}]{hestenes1952CG}%
  \BibitemOpen
  \bibfield  {author} {\bibinfo {author} {\bibfnamefont {M.~R.}\ \bibnamefont
  {Hestenes}}\ and\ \bibinfo {author} {\bibfnamefont {E.}~\bibnamefont
  {Stiefel}},\ }\href {http://dx.doi.org/10.6028/jres.049.044} {\bibfield
  {journal} {\bibinfo  {journal} {J. Res. Natl. Bur. Stan.}\ }\textbf {\bibinfo
  {volume} {49}},\ \bibinfo {pages} {409} (\bibinfo {year} {1952})}\BibitemShut
  {NoStop}%
\bibitem [{\citenamefont {Bradbury}\ and\ \citenamefont
  {Fletcher}(1966)}]{bradbury1966CGeigen}%
  \BibitemOpen
  \bibfield  {author} {\bibinfo {author} {\bibfnamefont {W.~W.}\ \bibnamefont
  {Bradbury}}\ and\ \bibinfo {author} {\bibfnamefont {R.}~\bibnamefont
  {Fletcher}},\ }\href {https://doi.org/10.1007/BF02162089} {\bibfield
  {journal} {\bibinfo  {journal} {Numer. Math.}\ }\textbf {\bibinfo {volume}
  {9}},\ \bibinfo {pages} {259} (\bibinfo {year} {1966})}\BibitemShut {NoStop}%
\bibitem [{\citenamefont {Knyazev}(1998)}]{Knyazev1998Preconditioned}%
  \BibitemOpen
  \bibfield  {author} {\bibinfo {author} {\bibfnamefont {A.~V.}\ \bibnamefont
  {Knyazev}},\ }\href@noop {} {\bibfield  {journal} {\bibinfo  {journal}
  {Electron. Trans. Numer. Anal.}\ }\textbf {\bibinfo {volume} {7}},\ \bibinfo
  {pages} {104} (\bibinfo {year} {1998})}\BibitemShut {NoStop}%
\bibitem [{\citenamefont {Knyazev}(2001)}]{Knyazev2001LOBPCG}%
  \BibitemOpen
  \bibfield  {author} {\bibinfo {author} {\bibfnamefont {A.~V.}\ \bibnamefont
  {Knyazev}},\ }\href {https://doi.org/10.1137/S1064827500366124} {\bibfield
  {journal} {\bibinfo  {journal} {SIAM J. Sci. Comput.}\ }\textbf {\bibinfo
  {volume} {23}},\ \bibinfo {pages} {517} (\bibinfo {year} {2001})}\BibitemShut
  {NoStop}%
\bibitem [{\citenamefont {Koepf}\ and\ \citenamefont
  {Ring}(1991)}]{koepf1991WoodsSaxon}%
  \BibitemOpen
  \bibfield  {author} {\bibinfo {author} {\bibfnamefont {W.}~\bibnamefont
  {Koepf}}\ and\ \bibinfo {author} {\bibfnamefont {P.}~\bibnamefont {Ring}},\
  }\href {http://dx.doi.org/10.1007/BF01282936} {\bibfield  {journal} {\bibinfo
   {journal} {Z. Phys. A}\ }\textbf {\bibinfo {volume} {339}},\ \bibinfo
  {pages} {81} (\bibinfo {year} {1991})}\BibitemShut {NoStop}%
\bibitem [{\citenamefont {Zhao}\ \emph {et~al.}(2015)\citenamefont {Zhao},
  \citenamefont {Itagaki},\ and\ \citenamefont {Meng}}]{Zhao2015Rod-shaped}%
  \BibitemOpen
  \bibfield  {author} {\bibinfo {author} {\bibfnamefont {P.~W.}\ \bibnamefont
  {Zhao}}, \bibinfo {author} {\bibfnamefont {N.}~\bibnamefont {Itagaki}}, \
  and\ \bibinfo {author} {\bibfnamefont {J.}~\bibnamefont {Meng}},\ }\href
  {\doibase 10.1103/PhysRevLett.115.022501} {\bibfield  {journal} {\bibinfo
  {journal} {Phys. Rev. Lett.}\ }\textbf {\bibinfo {volume} {115}},\ \bibinfo
  {pages} {022501} (\bibinfo {year} {2015})}\BibitemShut {NoStop}%
\bibitem [{\citenamefont {Zhao}\ \emph {et~al.}(2017)\citenamefont {Zhao},
  \citenamefont {Lu}, \citenamefont {Zhao},\ and\ \citenamefont
  {Zhou}}]{Zhao2017Phys.Rev.C14320}%
  \BibitemOpen
  \bibfield  {author} {\bibinfo {author} {\bibfnamefont {J.}~\bibnamefont
  {Zhao}}, \bibinfo {author} {\bibfnamefont {B.-N.}\ \bibnamefont {Lu}},
  \bibinfo {author} {\bibfnamefont {E.-G.}\ \bibnamefont {Zhao}}, \ and\
  \bibinfo {author} {\bibfnamefont {S.-G.}\ \bibnamefont {Zhou}},\ }\href
  {\doibase 10.1103/PhysRevC.95.014320} {\bibfield  {journal} {\bibinfo
  {journal} {Phys. Rev. C}\ }\textbf {\bibinfo {volume} {95}},\ \bibinfo
  {pages} {014320} (\bibinfo {year} {2017})}\BibitemShut {NoStop}%
\bibitem [{\citenamefont {Agbemava}\ \emph {et~al.}(2015)\citenamefont
  {Agbemava}, \citenamefont {Afanasjev}, \citenamefont {Nakatsukasa},\ and\
  \citenamefont {Ring}}]{Agbemava2015superheavy}%
  \BibitemOpen
  \bibfield  {author} {\bibinfo {author} {\bibfnamefont {S.~E.}\ \bibnamefont
  {Agbemava}}, \bibinfo {author} {\bibfnamefont {A.~V.}\ \bibnamefont
  {Afanasjev}}, \bibinfo {author} {\bibfnamefont {T.}~\bibnamefont
  {Nakatsukasa}}, \ and\ \bibinfo {author} {\bibfnamefont {P.}~\bibnamefont
  {Ring}},\ }\href {\doibase 10.1103/PhysRevC.92.054310} {\bibfield  {journal}
  {\bibinfo  {journal} {Phys. Rev. C}\ }\textbf {\bibinfo {volume} {92}},\
  \bibinfo {pages} {054310} (\bibinfo {year} {2015})}\BibitemShut {NoStop}%
\bibitem [{\citenamefont {Shi}\ \emph {et~al.}(2019)\citenamefont {Shi},
  \citenamefont {Afanasjev}, \citenamefont {Li},\ and\ \citenamefont
  {Meng}}]{Shi2019Superheavy}%
  \BibitemOpen
  \bibfield  {author} {\bibinfo {author} {\bibfnamefont {Z.}~\bibnamefont
  {Shi}}, \bibinfo {author} {\bibfnamefont {A.~V.}\ \bibnamefont {Afanasjev}},
  \bibinfo {author} {\bibfnamefont {Z.~P.}\ \bibnamefont {Li}}, \ and\ \bibinfo
  {author} {\bibfnamefont {J.}~\bibnamefont {Meng}},\ }\href {\doibase
  10.1103/PhysRevC.99.064316} {\bibfield  {journal} {\bibinfo  {journal} {Phys.
  Rev. C}\ }\textbf {\bibinfo {volume} {99}},\ \bibinfo {pages} {064316}
  (\bibinfo {year} {2019})}\BibitemShut {NoStop}%
\bibitem [{\citenamefont {Meng}\ \emph {et~al.}(2020)\citenamefont {Meng},
  \citenamefont {Lu},\ and\ \citenamefont {Zhou}}]{meng2020Hs270}%
  \BibitemOpen
  \bibfield  {author} {\bibinfo {author} {\bibfnamefont {X.}~\bibnamefont
  {Meng}}, \bibinfo {author} {\bibfnamefont {B.}~\bibnamefont {Lu}}, \ and\
  \bibinfo {author} {\bibfnamefont {S.}~\bibnamefont {Zhou}},\ }\href {\doibase
  https://doi.org/10.1007/s11433-019-9422-1} {\bibfield  {journal} {\bibinfo
  {journal} {Sci. China-Phys. Mech. Astron.}\ }\textbf {\bibinfo {volume}
  {63}},\ \bibinfo {pages} {212011} (\bibinfo {year} {2020})}\BibitemShut
  {NoStop}%
\bibitem [{\citenamefont {Lu}\ \emph {et~al.}(2012)\citenamefont {Lu},
  \citenamefont {Zhao},\ and\ \citenamefont {Zhou}}]{Lu2012Potential}%
  \BibitemOpen
  \bibfield  {author} {\bibinfo {author} {\bibfnamefont {B.-N.}\ \bibnamefont
  {Lu}}, \bibinfo {author} {\bibfnamefont {E.-G.}\ \bibnamefont {Zhao}}, \ and\
  \bibinfo {author} {\bibfnamefont {S.-G.}\ \bibnamefont {Zhou}},\ }\href
  {\doibase 10.1103/PhysRevC.85.011301} {\bibfield  {journal} {\bibinfo
  {journal} {Phys. Rev. C}\ }\textbf {\bibinfo {volume} {85}},\ \bibinfo
  {pages} {011301(R)} (\bibinfo {year} {2012})}\BibitemShut {NoStop}%
\bibitem [{\citenamefont {Lu}\ \emph {et~al.}(2014)\citenamefont {Lu},
  \citenamefont {Zhao}, \citenamefont {Zhao},\ and\ \citenamefont
  {Zhou}}]{Lu2014MCRMF}%
  \BibitemOpen
  \bibfield  {author} {\bibinfo {author} {\bibfnamefont {B.-N.}\ \bibnamefont
  {Lu}}, \bibinfo {author} {\bibfnamefont {J.}~\bibnamefont {Zhao}}, \bibinfo
  {author} {\bibfnamefont {E.-G.}\ \bibnamefont {Zhao}}, \ and\ \bibinfo
  {author} {\bibfnamefont {S.-G.}\ \bibnamefont {Zhou}},\ }\href {\doibase
  10.1103/PhysRevC.89.014323} {\bibfield  {journal} {\bibinfo  {journal} {Phys.
  Rev. C}\ }\textbf {\bibinfo {volume} {89}},\ \bibinfo {pages} {014323}
  (\bibinfo {year} {2014})}\BibitemShut {NoStop}%
\bibitem [{\citenamefont {Zhou}(2016)}]{Zhou2016PS}%
  \BibitemOpen
  \bibfield  {author} {\bibinfo {author} {\bibfnamefont {S.-G.}\ \bibnamefont
  {Zhou}},\ }\href {\doibase 10.1088/0031-8949/91/6/063008} {\bibfield
  {journal} {\bibinfo  {journal} {Phys. Scr.}\ }\textbf {\bibinfo {volume}
  {91}},\ \bibinfo {pages} {063008} (\bibinfo {year} {2016})}\BibitemShut
  {NoStop}%
\bibitem [{\citenamefont {Umar}\ and\ \citenamefont
  {Oberacker}(2015)}]{Umar2015SHE_TDHF}%
  \BibitemOpen
  \bibfield  {author} {\bibinfo {author} {\bibfnamefont {A.}~\bibnamefont
  {Umar}}\ and\ \bibinfo {author} {\bibfnamefont {V.}~\bibnamefont
  {Oberacker}},\ }\href {\doibase
  https://doi.org/10.1016/j.nuclphysa.2015.02.011} {\bibfield  {journal}
  {\bibinfo  {journal} {Nucl. Phys. A}\ }\textbf {\bibinfo {volume} {944}},\
  \bibinfo {pages} {238} (\bibinfo {year} {2015})}\BibitemShut {NoStop}%
\bibitem [{\citenamefont {Guo}\ \emph {et~al.}(2018)\citenamefont {Guo},
  \citenamefont {Shen}, \citenamefont {Yu},\ and\ \citenamefont
  {Wu}}]{Guo2018TDHF_fusion}%
  \BibitemOpen
  \bibfield  {author} {\bibinfo {author} {\bibfnamefont {L.}~\bibnamefont
  {Guo}}, \bibinfo {author} {\bibfnamefont {C.}~\bibnamefont {Shen}}, \bibinfo
  {author} {\bibfnamefont {C.}~\bibnamefont {Yu}}, \ and\ \bibinfo {author}
  {\bibfnamefont {Z.}~\bibnamefont {Wu}},\ }\href {\doibase
  10.1103/PhysRevC.98.064609} {\bibfield  {journal} {\bibinfo  {journal} {Phys.
  Rev. C}\ }\textbf {\bibinfo {volume} {98}},\ \bibinfo {pages} {064609}
  (\bibinfo {year} {2018})}\BibitemShut {NoStop}%
\end{thebibliography}


\end{document}